\documentclass[twocolumn]{aastex62}

\newcommand{\degree}{$^{\circ}$}

\newcommand{\hii}{H~{\scriptsize II}}

\received{\today}
\revised{\today}
\accepted{\today}

\submitjournal{ApJ}
\shorttitle{SNTF Polarimetric}
\shortauthors{Par\'e et al. 2024}

\begin{document}

\title{A VLA Study of Newly-Discovered Southern Latitude Non-Thermal Filaments in the Galactic Center: Polarimetric and Magnetic Field Properties}

\correspondingauthor{Dylan Par\'e}
\email{dylanpare@gmail.com}

\author[0000-0002-5811-0136]{Dylan M. Par\'e}
\affil{Villanova University \\
800 Lancaster Avenue \\
Villanova, PA 19085}

\author{Cornelia C. Lang}
\affil{University of Iowa \\
30 North Dubuque Street, Room 203 \\
Iowa City, IA 52242}

\author[0000-0002-6753-2066]{Mark R. Morris}
\affil{University of California, Los Angeles \\
430 Portola Plaza, Box 951547 \\
Los Angeles, CA 90095-1547}

\begin{abstract}
A population of structures unique to the Galactic Center (GC), known as the non-thermal filaments (NTFs), has been studied for over 40 years, but much remains unknown about them. In particular, there is no widely-accepted and unified understanding for how the relativistic electrons illuminating these structures are generated. One possibility is that there are compact and extended sources of Cosmic Rays (CRs), which then diffuse along magnetic flux tubes leading to the illumination of the NTFs through synchrotron emission. In this work, we present and discuss the polarimetric distributions associated with a set of faint NTFs in the GC that have only been studied in total intensity previously. We compare the derived polarized intensity, rotation measure, and intrinsic magnetic field distributions for these structures with the results obtained for previously observed GC NTFs. The results are then used to enhance our understanding of the large-scale polarimetric properties of the GC. We then use the derived polarimetric distributions to constrain models for the mechanisms generating the relativistic electrons that illuminate these structures.
\end{abstract}

\section{INTRODUCTION} \label{sec:intro}
The Galactic Center (GC) is by far the closest galactic nuclear region to Earth, being only $\rm\sim$8.2 kpc away \citep{Abuter2019,Do2019}. The GC displays extreme properties compared to those observed in the Galactic Disk, such as elevated molecular densities (10$\rm^3$-10$\rm^6$, \citealt{Mills2014,Mills2018}) and magnetic field strengths \citep[100s $\mu$G -- 10s mG, e.g.,][]{Yusef-Zadeh1987a,Plante1995,Chuss2003a,Pillai2015,Hsieh2018,Mangilli2019,Guerra2023}. Assuming the GC is a representative galactic nuclear region, we can use the GC to study the properties of nuclear regions that are too distant to be observed in detail with current instruments. 

A population of unique structures has been observed in the GC for the past few decades that appear as glowing threads (e.g. \citealt{YMC1984,Yusef-Zadeh1986b,YM1987,Gray1995,Lang1999a,Lang1999b,Pare2019,Pare2021,Pare2022,Yusef-Zadeh2022}). Because of the non-thermal emission from these structures, they have come to be known as the non-thermal filaments (NTFs). The NTFs are highly polarized sources (with percentage polarizations $\rm\sim$50\%) that are illuminated by synchrotron emission from relativistic electrons \citep{Morris1996}. However, the mechanism for accelerating the relativistic electrons remains unclear \citep[see][and references therein]{Ponti2021}.

The first observed NTF, known as the Radio Arc, is also the most prominent of these structures \citep{YMC1984,Yusef-Zadeh1986a,YM1987}. The Radio Arc consists of $>$10 individual filaments, with each filament having a length of $\sim$20\arcmin\ ($\sim$45 pc) and a narrow width of 0.5\arcsec\ \citep[0.02 pc,][]{Pare2019}.

Until recently, the Radio Arc was the only NTF known to possess so many individual filaments (e.g. a ``multi-stranded'' filament or ``filament bundle''). By comparison, most other NTFs appear to be comprised of only one or a few filaments (``single-stranded'' NTFs), such as the Snake and Pelican NTFs and Northern and Southern threads \citep{Morris1985,Gray1995,Lang1999a,Lang1999b}. However, observations at 1 GHz using MeerKAT have revealed a population of faint NTFs that appear to be morphologically similar to the Radio Arc in that they each consist of $\rm\geq$10 individual filaments \citep{Thomas2020,Heywood2022,Yusef-Zadeh2022}. These newly-discovered NTF bundles (NTFBs) are in more isolated regions of the GC than the Radio Arc and are located at Southern latitudes of the GC. The Southern latitudes of these structures contrasts with the predominantly Northern latitudes of most previously observed NTFs \citep[e.g.,][]{Gray1995,Lang1999a,Lang1999b}. 

The existence of these additional multi-stranded NTFs makes it possible to explore mechanisms for how the relativistic electrons illuminating the NTFs are generated. These electrons could be generated through various processes like magnetic reconnection off the surface of GC molecular clouds \citep{Serabyn1994}, diffusion from pulsar wind nebulae \citep{Thomas2020}, Cosmic Ray (CR) propagation originating from large-scale high-energy X-ray structures in the region \citep{Ponti2021}, or diffusive shock acceleration in shock fronts from stellar winds and ionization fronts \citep{Rosner1996}. Because of the variation in NTF morphology and their surrounding environments, a unified model that explains the origin of the relativistic electrons illuminating all of the GC NTFs has remained elusive.

The total intensity properties of a selection of faint, multi-stranded NTFs were recently analyzed and it was determined that there are multiple possible mechanisms that could be responsible for illuminating the NTFBs \citep{Pare2022}. Some of the NTFBs show indications of being illuminated by a compact source of CRs, such as the harps studied by \citet{Thomas2020} and SNTF1 studied by \citet{Pare2022}, whereas others manifest properties indicative of being illuminated by an extended source of CRs, such as the Radio Arc, Snake, and SNTF3 \citep{Yusef-Zadeh1987,Gray1995,Pare2022}. There is other observational evidence for both compact and extended sources of CRs in the GC, like pulsar wind nebulae and large-scale outflow structures detected over multiple wavelengths \citep[radio, infrared, and x-ray,][]{Ponti2019,Heywood2019,Thomas2020,Ponti2021}.

An important aspect of the investigation of the NTFBs that is missing in \citet{Pare2022} is the polarimetric properties of these structures. Analysis of the polarized intensity, rotation measure (RM), and intrinsic magnetic field distributions of these structures can be used to further inform our understanding of how the NTFBs are powered. Such an analysis could also inform our understanding of how the larger GC NTF population is illuminated through comparisons with the properties observed for the Radio Arc and other GC NTFs.

\begin{figure*}
    \centering
    \includegraphics[width=1.0\textwidth]{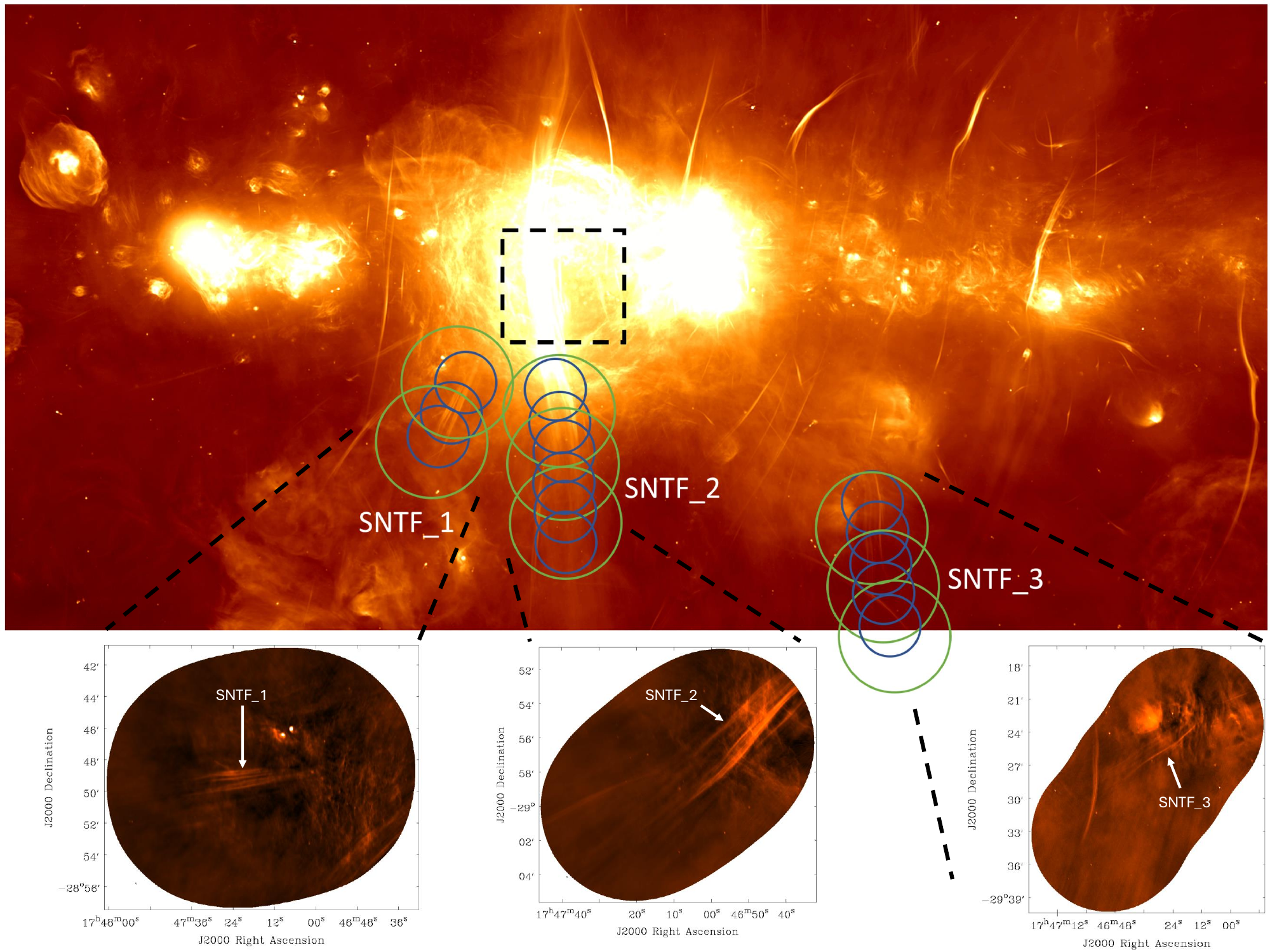}
    \caption{1 GHz MeerKAT image of the GC \citep{Heywood2022} with the VLA fields of view for our NTFB observations overlayed as circles. 10 GHz fields of view are in blue and 6 GHz fields of view are in green. The black rectangle indicates the field of view of other previous Radio Arc observations at 6 and 10 GHz \citep{Pare2019,Pare2021}. Inset panels show the 6 GHz total intensity distributions for the NTFBs observed in \citet{Pare2022}.}
    \label{fig:legend} 
\end{figure*}
In this paper we build on the work presented in \citet{Pare2022} by studying the polarimetric, RM, and intrinsic magnetic field distributions of the same NTFBs observed in that work. The locations of these NTFBs are marked and labelled in Figure \ref{fig:legend} where we use the same naming convention for the NTFB targets as used in \citet{Pare2022}. As shown in Figure \ref{fig:legend} the NTFBs were observed using multiple fields of view with the Karl G. Jansky Very Large Array (VLA), managed by the NRAO,\footnote{The National Radio Astronomy Observatory is a facility of the National Science Foundation operated under cooperative agreement by Associated Universities, Inc.} at both C- and X-bands (central frequencies of 6 and 10 GHz respectively) as indicated with the green and blue circles in Figure \ref{fig:legend}. Inset panels of Figure \ref{fig:legend} show the 6 GHz total intensity distributions for the NTFBs studied in \citet{Pare2022}.

In Section \ref{sec:dat_red} we detail the data calibration and imaging methods used to generate the polarimetric data presented in this paper. In Section \ref{sec:P_res} we present the 3$\rm\sigma$ debiased polarized intensity distributions for our target NTFBs. In Section \ref{sec:RM_res} we present the RM distributions obtained for our targets, and in Section \ref{sec:B_res} we present the intrinsic magnetic field distributions, corrected for intervening Faraday effects, for our targets. In Section \ref{sec:disc} we discuss the key interpretations of our results. We conclude the paper in Section \ref{sec:conc}.

\section{OBSERVATIONS AND DATA REDUCTION} \label{sec:dat_red}

\subsection{Observations} \label{sec:obs}
\begin{deluxetable*}{|c|c|c|c|c|c|c|}[ht!]
\tablecaption{Summary of NTFB VLA Observations}
\tablecolumns{7}
\tablenum{1}
\tablewidth{0pt}
\tablehead{
\colhead{Object} & \colhead{Freq. Band} & \colhead{Freq. Range (GHz)} & \colhead{T$\rm_{int}$/field (min)} & \colhead{N$\rm_{fields}$} & \colhead{Array Configs.} & \colhead{Total Time (hrs)}
}
\startdata
SNTF1 & C & 4.0 - 8.0 & 50 & 2 & B, C & 1.67 \\
      & X & 8.0 - 12.0 & 45 & 3 & C & 1.5 \\
SNTF2 & C & 4.0 - 8.0 & 50 & 3 & B, C & 2.5 \\
      & X & 8.0 - 12.0 & 45 & 6$^*$ & C & 3.0 \\
SNTF3 & C & 4.0 - 8.0 & 50 & 3 & B, C & 2.5 \\
      & X & 8.0 - 12.0 & 45 & 5 & C & 2.50 \\
\enddata
\tablecomments{``Object'' indicates the name of the observed NTFB, ``Freq. band'' denotes the name of the frequency band used for the observation, ``Freq. Range'' shows the frequency range of the frequency band, ``T$\rm_{int}$/field'' shows the amount of time observed for each pointing, ``N$\rm_{fields}$'' shows the number of fields observed, ``Array Configs''. details the VLA arrays used for each frequency band, and ``Total Time'' shows the total observation time in each frequency band. The $^*$ for SNTF3 $N_{fields}$ indicates that the two Southernmost X-band fields of SNTF3 were not actually observed due to observer error.}
\end{deluxetable*}

The observations used for this paper are the same observations as those presented by \citet{Pare2022}. Three NTFBs were observed at both C- and X-bands (at frequencies of 6 and 10 GHz respectively) using the B- and C-configurations of the VLA during the Spring of 2020. Multiple VLA pointings were used for each target to fully image each NTFB at the two frequency bands observed, as shown in Figure \ref{fig:legend}. The details of the observations are shown in Table 1. Though bright compact sources were located near the edge of the primary beam in some instances there were no substantial artifacts produced by these edge sources, as described in \citet{Pare2022}.

\subsection{Calibration} \label{sec:cal}

Three calibrators were observed at each frequency band. 3C 286 was used as the flux and bandpass calibrator, with J1407 used as the polarization calibrator and 1751-253 used as the phase calibrator. 3C 286 and J1407 were observed once at the beginning of each observation for each of the NTFBs, whereas 1751-253 was observed every 8 minutes during the observations of the targets.

To calibrate the polarimetric observations presented in this paper, we began with the total intensity calibrated observations used to produce the results presented in \citet{Pare2022}. The observations in \citet{Pare2022} were calibrated using the Common Astronomy Software Applications (CASA) VLA Calibration Pipeline\footnote{https://science.nrao.edu/facilities/vla/data-processing/pipeline}; however, this pipeline does not currently support polarization calibration. Therefore, additional calibration steps were needed to obtain finalized polarimetric data sets. 

The additional polarimetric calibration was performed on the total intensity calibrated measurement sets following the guidelines in the NRAO CASA guides (see CASA guides on calibrating 3C75 for CASA version 6.2.1\footnote{https://casaguides.nrao.edu}). These additional calibration steps involve the correction of cross-hand delays, polarization leakage, and polarization angle differences. These polarimetric corrections were implemented using the CASA tasks \textit{gaincal()} and \textit{polcal()}. We obtained well calibrated observations by using the basic calibration parameters for these steps, and no additional calibration steps or considerations were needed for these observations.

\subsection{Imaging} \label{sec:imag}
Two sets of images were produced from these calibrated observations: \textbf{1)} 2D polarized intensity distributions to analyze the polarized intensity morphology and \textbf{2)} polarization spectral image cubes to derive the RM and intrinsic magnetic field distributions. 

\subsubsection{2D Polarized Intensity Distributions} \label{sec:p_2d}
Polarimetric observations from the VLA are comprised of the individual Stokes parameters: $I$, $Q$, $U$, and $V$, where Stokes $I$ corresponds to the total intensity emission of the target, Stokes $Q$ and $U$ are the linear polarization products, and Stokes $V$ corresponds to circular polarization. The Stokes $I$ total intensity distributions of the NTFBs were studied in \citet{Pare2022}, examples of which are shown in the inset panels of Figure \ref{fig:legend}; in this work we focus on the $Q$, $U$, and $V$ distributions of these targets. 

We first inspected the Stokes $V$ circular polarization distributions and found no significant circular polarization for any of our NTFB targets. Circular polarization has not been observed for other GC NTF targets, so the lack of significant Stokes $V$ emission is expected. We do, however, obtain significant Stokes $Q$ and $U$ linear polarization emission.

To analyze the linear Stokes $Q$ and $U$ products we used the CASA \textit{tclean} task in mfs mode with 100 iterations to produce separate Stokes $Q$ and $U$ data sets. The combination of the Stokes $Q$ and $U$ linear polarization emission allows us to derive the polarized intensity and observed polarization angle distributions for our targets using the following equations:
\begin{equation}
    P = \sqrt{Q^2+U^2} \label{eq:p}
\end{equation} and
\begin{equation}
    \chi = \frac{1}{2}tan^{-1}\left(\frac{U}{Q}\right) \label{eq:chi},
\end{equation}
where $P$ (Jy beam$^{-1}$) is the polarized intensity and $\chi$ (rad) is the observed polarization angle. The calculations shown in Equations \ref{eq:p} and \ref{eq:chi} are performed using the CASA task \textit{immath} on the cleaned $Q$ and $U$ (Jy beam$^{-1}$) observations. 

After obtaining the polarized intensity distribution we then debiased the polarization. Debiasing helps account for artifacts in the polarization resulting from decreased polarization sensitivity near the edge of the VLA primary beam and other instrumental effects. The polarized intensity distributions analyzed in this work have been debiased at the 3$\sigma_P$ level such that:
\begin{equation}
    P_d = P - 3\times\sigma_P, \label{eq:PD}
\end{equation}
where $P_d$ (Jy beam$^{-1}$) is the debiased polarized intensity, $P$ is the polarized intensity derived from the cleaned $Q$ and $U$ distributions as shown in Equation \ref{eq:p}, and $\sigma_P$ (Jy beam$^{-1}$) is the rms noise of the polarized intensity distribution.

The 3$\sigma_P$ debiasing is the last step performed to obtain scientifically robust polarized intensity distributions for our targets.
These 2D, debiased polarized intensity distributions are presented in Section \ref{sec:P_res}. Regions of polarization that are removed as a result of debiasing are due to low SNR and not low fractional polarization.

\subsubsection{Spectral Image Cubes}
To derive the intrinsic magnetic field distributions of the NTFBs an understanding of how the polarization of our targets varies as a function of frequency is needed. To achieve that we created a second set of images that are image cubes with axes of position (Right Ascension), position (Declination), and frequency. We made these cubes using the CASA task \textit{tclean}, performing a separate mfs-mode clean on each spectral window (spw) of our observations. For each spw we obtained separate Stokes $I$, $Q$, and $U$ distributions. We did not produce spectral cubes for the Stokes $V$ parameter since no significant circular polarization was observed in the 2D polarimetric distributions. These individual spw distributions were then inspected for the presence of artifacts and elevated noise. Any spws that were found to possess significant artifacts or an rms noise more than 2$\times$ above the average noise level of the set of spws were ignored in creating the final spectral image cubes.

The remaining spws were then smoothed to a common restoring beam size corresponding to the largest beam size found in the set of spws for a particular NTFB. These smoothed spws were then combined in frequency space into an image cube along the spectral axis, resulting in 3D image cubes of Stokes $I$, $Q$, and $U$ for each NTFB. Table 2 indicates which spws were used to make our cubes for each NTFB, as well as the number of spws in the final image cubes. 

Though we made image cubes for all of our 6 and 10 GHz observations we did not detect significant polarization at 6 GHz for SNTF1, SNTF2, or SNTF3. We also did not detect significant polarization at 10 GHz for SNTF2. The lack of significant polarization is due to low signal to noise for these observations, rather than these filaments having unexpectedly low fractional polarizations. We find that the fractional polarizations for the polarizations that are present in the filamentary structures before debiasing have fractional polarizations ranging from 0.1 to 0.5, characteristic fractional polarizations for GC NTFs.

The Wishbone NTF was also observed as part of the SNTF3 observations. This filament is covered in the 6 GHz observations and significant polarization is obtained from the Wishbone from these observations. Table 2 only provides information for the data cubes containing significant polarization, as we chose to focus on the significant detections obtained from our observations in this work.

For the spectral cubes with significant polarization we then ran the CASA task \textit{immath} on the final Stokes $Q$ and $U$ image cubes to generate spectral cubes of polarized intensity and polarization angle, in a manner analogous to the procedure described in Section \ref{sec:p_2d}. These cubes were then used to derive the rotation measure (RM) and intrinsic magnetic field distributions. The resulting RM distributions are shown in Section \ref{sec:RM_res} and the intrinsic magnetic field distributions are shown in Section \ref{sec:B_res}. The methods used to produce these distributions using the spectral cubes are also presented in these sections.

\begin{deluxetable}{|c|c|c|c|}[ht!]
\tablecaption{Composition of Spectral Cubes for Each NTFB}
\tablecolumns{4}
\tablenum{2}
\tablewidth{0pt}
\tablehead{
\colhead{Object} & \colhead{Freq. Band} & \colhead{spws} & \colhead{N$\rm_{spw}$}
}
\startdata
SNTF1 & X & 0--31 & 32 \\
SNTF3 & X & 0--28 & 29\\
Wish  & C & 0--28 & 29 \\
\enddata
\tablecomments{Object indicates the name of the NTFB observed, Freq. band denotes the name of the frequency band used for the observation, spws indicates which spectral windows were used to make the cube (0-indexed, with increasing number indicating higher central frequency), and N$\rm_{spw}$ indicates the number of spws comprising the cube.}
\end{deluxetable}

\section{DEBIASED POLARIZED INTENSITY DISTRIBUTIONS} \label{sec:P_res}

\subsection{SNTF1} \label{sec{SNTF1_p}}
\begin{figure*}
    \centering
    \includegraphics[width=1.0\textwidth]{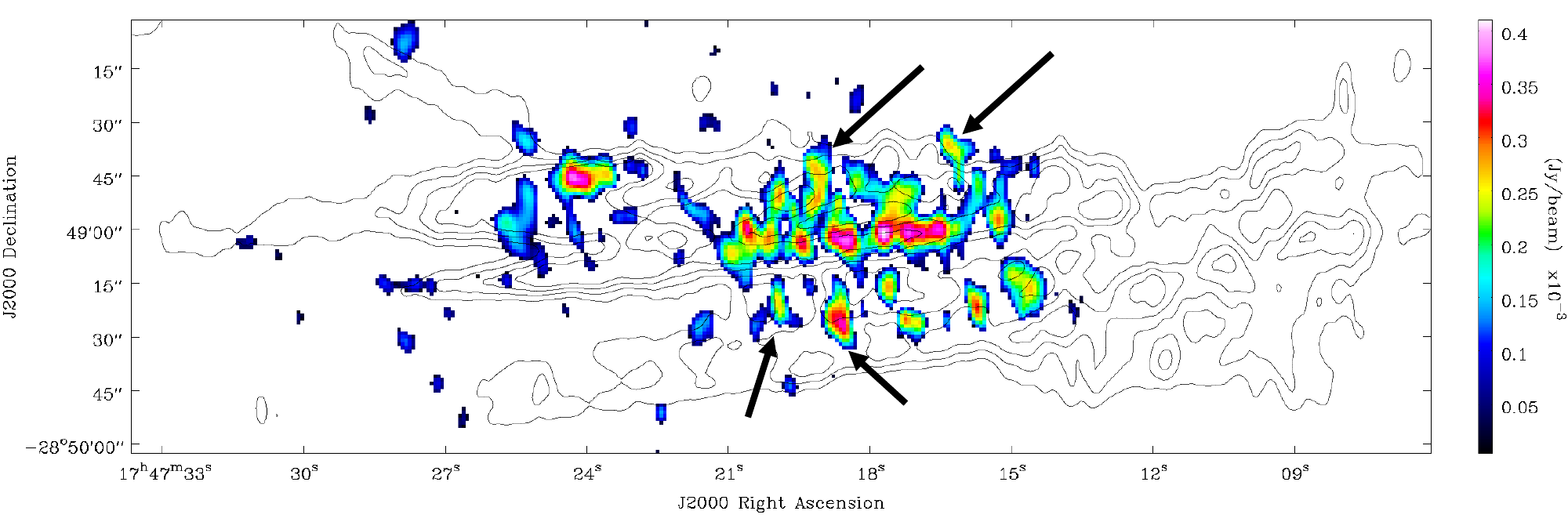}
    \caption{Debiased polarized intensity (in color) for SNTF1 at 10 GHz with polarization less than 3$\sigma_P$ above the noise level (52.5 $\mu$Jy beam$^{-1}$) masked out. Contours show the 10 GHz total intensity with levels of 0.02, 0.04, 0.06, and 0.08 $\times$ the peak total intensity of 5.9 mJy beam$^{-1}$. Both the polarized and total intensity data sets have been smoothed to the beam-size used to determine the RM and magnetic field distributions (6.0\arcsec $\times$ 2.4\arcsec, pa=2.4\degree), with the beam shown in the lower-left corner of the image. Black arrows indicate some of the polarized intensity extensions discussed in the text.}
    \label{fig:SNTF1P_2D}
\end{figure*}
We show the 2D debiased polarized intensity for SNTF1 in Figure \ref{fig:SNTF1P_2D}. The polarized intensity is most significant in a concentrated central region of the SNTF1 filaments ranging from roughly RA = 17:47:26 to RA = 17:47:14. This RA range generally corresponds to the region where the filaments of SNTF1 are brightest in total intensity \citep{Pare2022}. 

The polarized intensity appears ``patchy'' or discontinuous, which is a similar polarized intensity morphology to what is observed for the other NTFs in the GC \citep[e.g.,][]{Morris1996,Lang1999b}. We do not see significant polarization coinciding with the "jetted point source" identified in \citet{Pare2022} and visible in the contours of Figure \ref{fig:SNTF1P_2D} at RA = 17:47:28, DEC = -28:48:15. Notably, there appears to be an enhancement in polarized intensity where the jet or tail of this point source approaches the filaments of SNTF1. This enhancement is visible at RA = 17:47:24, DEC = -28:48:45 in Figure \ref{fig:SNTF1P_2D}.

We also study the fractional polarization $p$/$I$ for SNTF1, finding an average fractional polarization of 0.1 for significant polarization coinciding with the SNTF1 filaments. The fractional polarization is generally constant throughout the NTFB structure. The average fractional polarization of 0.1 is lower than what is observed for the majority of GC NTFs which are observed to have larger fractional polarizations of $>$0.3 \citep{YWP1997,Lang1999a,Lang1999b}.

We observe extensions of the polarized intensity that do not coincide with significant total intensity emission. Examples of these extensions are marked with black arrows in Figure \ref{fig:SNTF1P_2D}. In these polarization extensions we observe fractional polarizations $>$1.0 due to the lack of a significant total intensity counterpart. These structures and their large $p$/$I$ values are discussed in more detail below in Section \ref{sec:disc}.

\subsection{SNTF2 \& the Quill} \label{sec:SNTF2_p}
Our SNTF2 pointings encompass both the SNTF2 NTFB complex and a fainter NTF known as the Quill \citep{Pare2022}. The mosaic distributions combining all of the individual pointings at the different frequencies resulted in no significant polarization. To attempt to extract a significant detection from the Quill and SNTF2 separately we used \textit{tclean} to clean the individual 6 and 10 GHz pointings (the individual circles covering SNTF2 shown in Figure \ref{fig:legend}). Doing so results in separate fields-of-view along the length of SNTF2. Since the intensity of SNTF2 decreases local to the Quill, the separate fields-of-view ideally provide improved insight into the polarimetric properties of the fainter Quill NTF.

However, even with imaging the separate pointings we do not recover significant polarized emission from either SNTF2 or the Quill, likely because of how faint these structures are compared to the rms noise level of the observations. As a result, we do not display any polarized intensity results for either SNTF2 or the Quill in this paper, choosing to prioritize the significant detections obtained for our other NTFB targets. We note, however, that the fractional polarizations obtained from the un-debiased polarized intensity distributions of these structures yield fractional polarizations of $\sim$0.3, indicating that the lack of significant detections is a result of low SNR. 

\subsection{SNTF3 \& the Wishbone} \label{sec:SNTF3_p}
\begin{figure}
    \centering
    \includegraphics[scale=0.4]{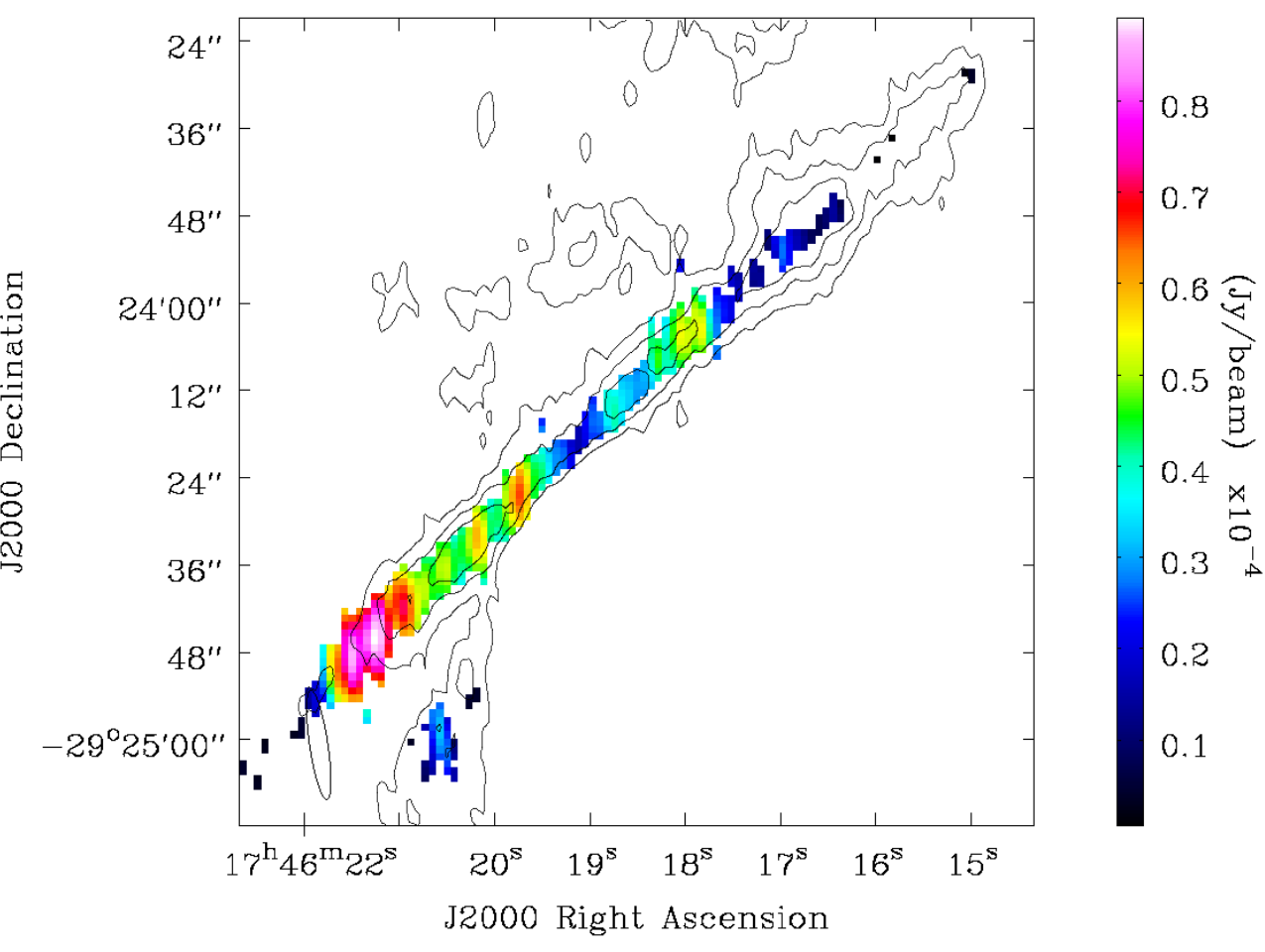}
    \caption{Debiased polarized intensity (in color) for SNTF3 at 10 GHz with polarization less than 3$\rm\sigma_P$ above the noise level (30.1 $\mu$Jy beam$^{-1}$) masked out. Contours show the 10 GHz total intensity with levels of 0.2, 0.3, 0.4, and 0.6 $\times$ the peak total intensity of 0.3 mJy beam$^{-1}$. The polarized intensity has been smoothed to the beam-size used to determine the RM and magnetic field distributions (15.2\arcsec $\times$ 2.5\arcsec, pa=8.5\degree), which is shown in the lower-left corner of the image.}
    \label{fig:SNTF3_X_P_2D}
\end{figure}
The 10 GHz debiased polarized intensity for SNTF3 is shown in Figure \ref{fig:SNTF3_X_P_2D}. The polarized intensity is very contiguous, which contrasts with the discontinuous or ``patchy'' polarized intensity distribution seen for SNTF1 and other previously observed GC NTFs \citep[e.g.][]{Pare2019,Lang1999a,Lang1999b,Gray1995}. The peak polarized intensity is seen at the Southern end of the brightest SNTF3 filament, located at RA = 17:46:21, DEC = -29:25:48, with the polarized intensity decreasing towards the Northern portion of the filament. Inspecting the fractional polarization we find the Southern portion of the NTF filament exhibits average fractional polarizations of 0.4. The fractional polarization progressively decreases further North along the filament length, decreasing to an average fractional polarization of 0.1. The higher fractional polarization regime in the South of SNTF3 is in good agreement with the fractional polarizations observed in the larger NTF population of $>$0.3 \citep{YWP1997,Lang1999a,Lang1999b}. The lower fractional polarizations to the North of SNTF3 are lower than generally observed for the larger NTF population, but are similar to the average fractional polarization observed for SNTF1.

\begin{figure}
    \centering
    \includegraphics[width=0.41\textwidth]{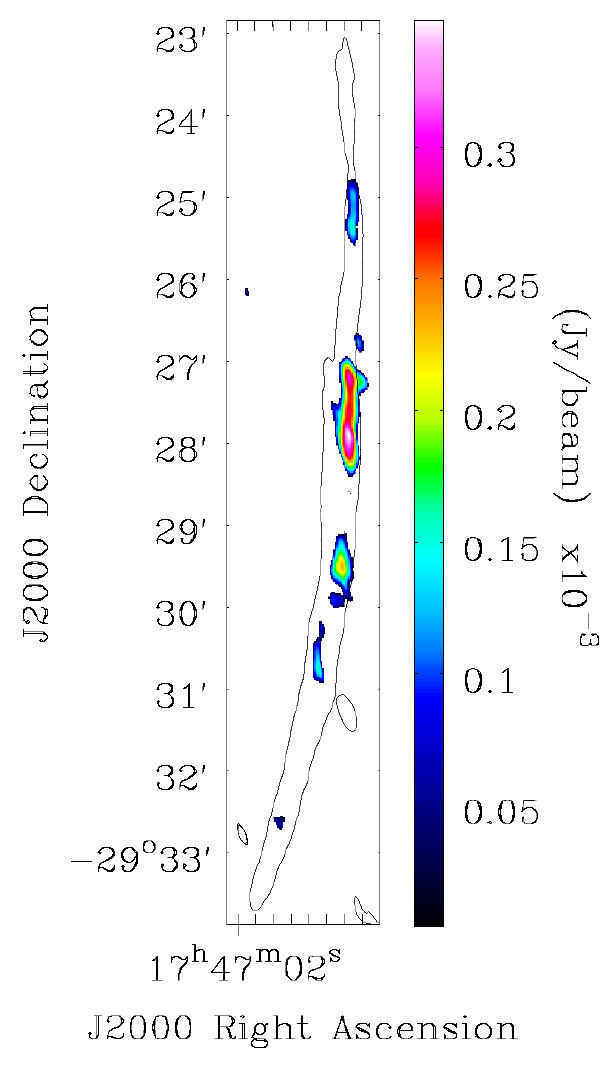}
    \caption{Debiased polarized intensity distribution (in color) for the Wishbone at 6 GHz with polarization less than 3$\sigma_P$ above the noise level (200 $\mu$Jy beam$^{-1}$) masked out. Contour level is of the 6 GHz total intensity at 0.02 $\times$ the peak total intensity value of 3.89 mJy beam$\rm^{-1}$. The polarized intensity has been smoothed to the beam-size used to determine the RM and intrinsic magnetic field distributions (16.7\arcsec\ $\times$ 4.9\arcsec, pa=22\degree), which is shown in the lower-left corner of the image.}
    \label{fig:Wish_C_P_2D}
\end{figure}

Figure \ref{fig:Wish_C_P_2D} shows the 6 GHz debiased polarized intensity distribution for the Wishbone NTF, which is close in projection to SNTF3 \citep{Pare2022}. This NTF was not covered by the field of view of the 10 GHz observations, so we only obtain polarimetric distributions derived from our 6 GHz observations. We note that the observations were constructed to include two 10 GHz pointings that covered the Wishbone, but these fields were not observed due to observer error. This NTF is unusual because we obtain significant polarized intensity at 6 GHz, whereas our other NTFB targets are fully depolarized at this lower frequency range. The polarized intensity for the Wishbone is quite discontinuous, but regions of significant polarization are roughly evenly spaced throughout the length of the NTF filament. The peak polarized intensity is seen near the middle of the Wishbone filament along its length at RA = 17:46:55, DEC = -29:27:30.

The fractional polarization for the Wishbone varies throughout its length, but the average fractional polarization is roughly 0.4. The fractional polarization is generally higher in the Northern portion of the filament (with values ranging from 0.45 - 0.5), and lower in the Southern portion (with value ranging from 0.25 - 0.35). These fractional polarizations throughout the filament length are in general agreement with what is observed in the larger NTF population \citep{YWP1997,Lang1999a,Lang1999b}.

\section{RM Results} \label{sec:RM_res}
\subsection{Derivation of RM} \label{sec:RM_der}
The spectral cubes of polarization angle were used to produce RM distributions for the SNTFs. The RM was derived by fitting a linear model to the polarization angle as a function of wavelength squared:
\begin{equation}
    \rm \chi = RM\times\lambda^2 + \chi_0 \label{eq:RM}
\end{equation}
where $\rm\chi$ (rad) is the observed polarization angle, RM (rad m$\rm^{-2}$) is the rotation measure encountered along the line of sight, $\rm\lambda$ (m) is the wavelength of the observation, and $\rm\chi_0$ (rad) is the intrinsic polarization angle at the source of the polarized emission. By fitting a linear model to $\rm\chi$ vs $\rm\lambda^2$ one can obtain the RM from the slope of the model and $\rm\chi_0$ from the y-intercept of the model. RM values were only derived for lines of sight coinciding with significant polarized intensity as shown in the figures in Section \ref{sec:P_res} (Figures \ref{fig:SNTF1P_2D} -- \ref{fig:Wish_C_P_2D}). 

\subsection{SNTF1} \label{SNTF1_RM}
\begin{figure*}
    \centering
    \includegraphics[width=1.0\textwidth]{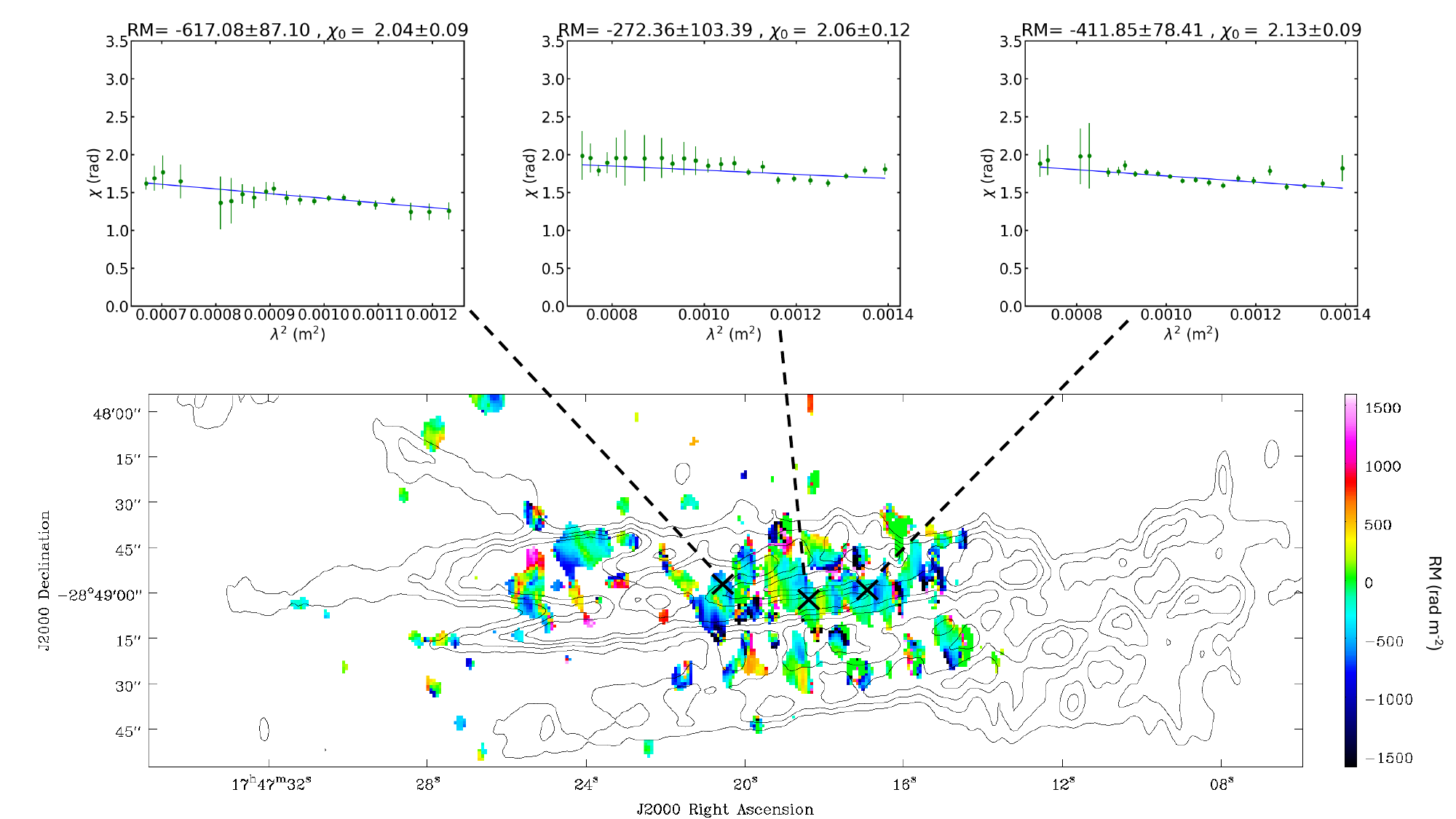}
    \caption{Color representation of the RM distribution in SNTF1, as obtained from the 10 GHz observations. The black contours represent the same total intensity levels as shown in Figure \ref{fig:SNTF1P_2D}. Inset panels show example fits of the polarization angle (green data points) as a function of $\rm\lambda^2$ with the best-fit lines shown in blue.}
    \label{fig:SNTF1_RM}
\end{figure*}
The RM distribution obtained from a least-squares fitting for SNTF1 is shown in Figure \ref{fig:SNTF1_RM}, with representative fits also displayed. Polarization angle values having an error greater than 0.5 rad are ignored in the least-squares fitting. The high errors that are discarded are a result of spws that have higher noise levels in Stokes $Q$ and $U$ at those spatial locations.

Figure \ref{fig:SNTF1_RM} reveals a generally flat RM distribution for SNTF1, with RM magnitudes generally $\rm|RM|<$1000 rad m$^{-2}$. Larger RMs approaching $\rm|RM|>$1500 rad m$^{-2}$ are observed toward several small patches in SNTF1. We note that the RM uncertainties in these regions are generally $\sim$400 rad m$^{-2}$, meaning these uncertainties make the RMs in these patches consistent with a distribution that is $<$1000 rad m$^{-2}$. These regions have the largest RM uncertainties with more general uncertainties being $\sim$100 rad m$^{-2}$. There is no significant RM gradient in the polarized portion of SNTF1.

\subsection{SNTF3 \& the Wishbone} \label{sec:SNTF3_RM}
\begin{figure*}
    \centering
    \includegraphics[width=1.0\textwidth]{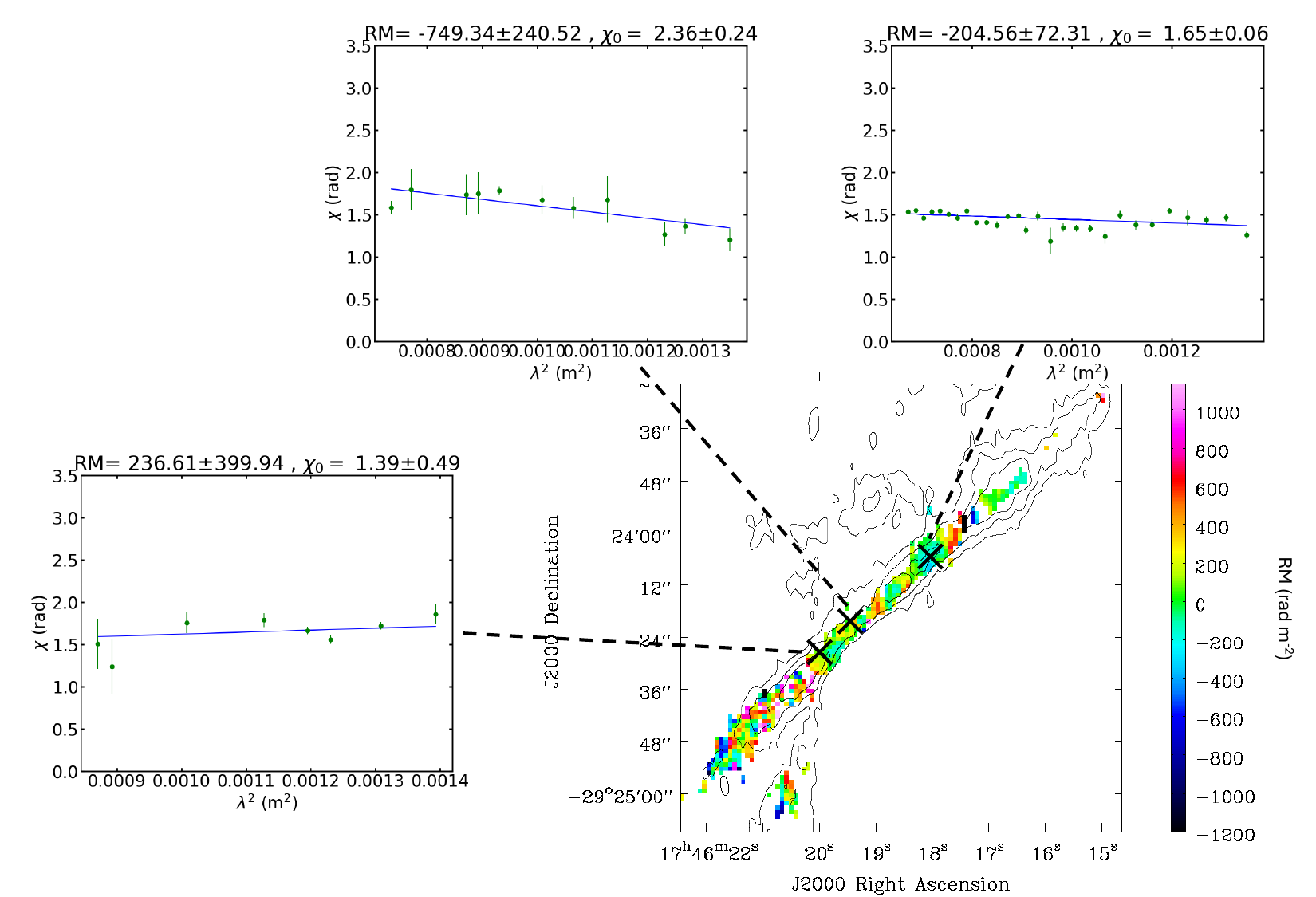}
    \caption{Color representation of the RM distribution obtained for SNTF3 at 10 GHz. The black contours represent the same total intensity levels as in Figure \ref{fig:SNTF3_X_P_2D}. Inset panels show example fits of the polarization angle (green data points) as a function of $\rm\lambda^2$ with the best-fit lines shown in blue.}
    \label{fig:SNTF3_X_RM}
\end{figure*}
The RM distribution obtained for SNTF3 from our 10 GHz observations is shown in Figure \ref{fig:SNTF3_X_RM}. The RM magnitudes for this filament are generally $\rm|RM|<$500 rad m$\rm^{-2}$. There is an apparent RM gradient seen for this NTFB where the Southern portion of SNTF3 has generally larger RM magnitudes of $|RM|>$600 rad m$^{-2}$, compared to the Northern portion of SNTF3. However, the RM values observed in this region are more stochastic which could be a result of lower SNR in this region. This possibility is particularly relevant since the total intensity emission from SNTF3 is fainter in this region.

We do not obtain significant polarized intensity for SNTF3 at 6 GHz although we do for the nearby Wishbone NTF structure that was also studied in \citet{Pare2022}. However, because of the lower SNR of our polarization detection at 6 GHz for the Wishbone we do not obtain high quality spectral polarimetric data. We therefore do not present the RM or intrinsic magnetic field distirubtions for the Wishbone because of the high uncertainties in our model fit parameters, especially since the polarized intensity distribution for the Wishbone at 6 GHz is very sparse.

\section{Intrinsic Magnetic Field Distributions} \label{sec:B_res}
\subsection{Derivation of the Intrinsic Magnetic Field} \label{sec:B_der}
Using the RM values obtained from our linear fits shown in Section \ref{sec:RM_res} it is possible to correct for the intervening Faraday effects encountered along the line of sight and thereby determine the intrinsic magnetic field orientation, $\rm\chi_0$, in the plane of the sky at the source of the polarized emission:
\begin{equation}
    \chi_0 = \chi - RM\times\lambda^2. \label{eq:chi0}
\end{equation}

The intrinsic polarization angle obtained from this equation is then rotated by 90\degree\ to obtain the intrinsic orientation of the plane-of-sky component of the magnetic field at the source of the polarized emission.

\subsection{SNTF1} \label{sec:SNTF1_B}
\begin{figure*}
    \centering
    \includegraphics[width=1.0\textwidth]{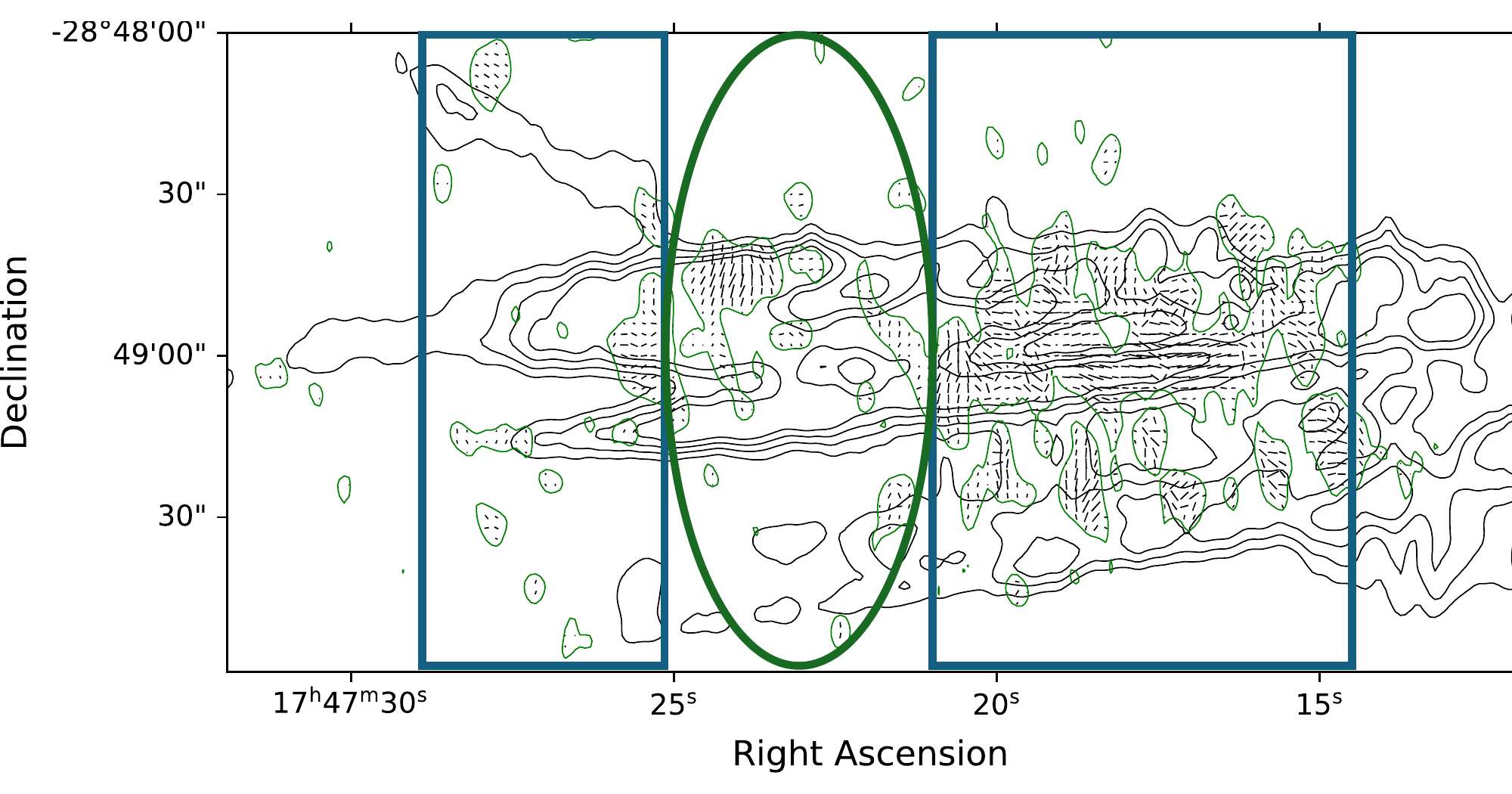}
    \caption{Intrinsic magnetic field orientations derived from the 10 GHz observations of SNTF1. Black contours indicate levels of constant total intensity at 0.15, 0.25, 0.35, and 0.45 mJy beam$\rm^{-1}$. Green elliptical and blue rectangular boxes mark zones where the magnetic field in SNTF1 is respectively predominantly perpendicular and parallel to the orientation of the total intensity filaments.}
    \label{fig:SNTF1_X_B}
\end{figure*}
Figure \ref{fig:SNTF1_X_B} shows the intrinsic magnetic field orientations for SNTF1 derived from regions of significant polarized intensity and using the RM values displayed in Figure \ref{fig:SNTF1_RM}. The intrinsic magnetic field of SNTF1 is spatially variant, with magnetic field orientations ranging from being parallel to perpendicular with respect to the NTF filament orientation. There is a slight alternating pattern in the magnetic field orientation which is marked in Figure \ref{fig:SNTF1_X_B}, with blue rectangular regions indicating zones of predominantly parallel magnetic field orientation. The green elliptical region marks a zone of predominantly perpendicular magnetic field orientation.

In addition to this alternating pattern along the filament lengths, there is a general trend of more perpendicular magnetic field orientation in Southern declination SNTF1 filaments compared to the Northern declination SNTF1 filaments. 

\subsection{SNTF3} \label{sec:SNTF3_B}
\begin{figure*}
    \centering
    \includegraphics[width=1.0\textwidth]{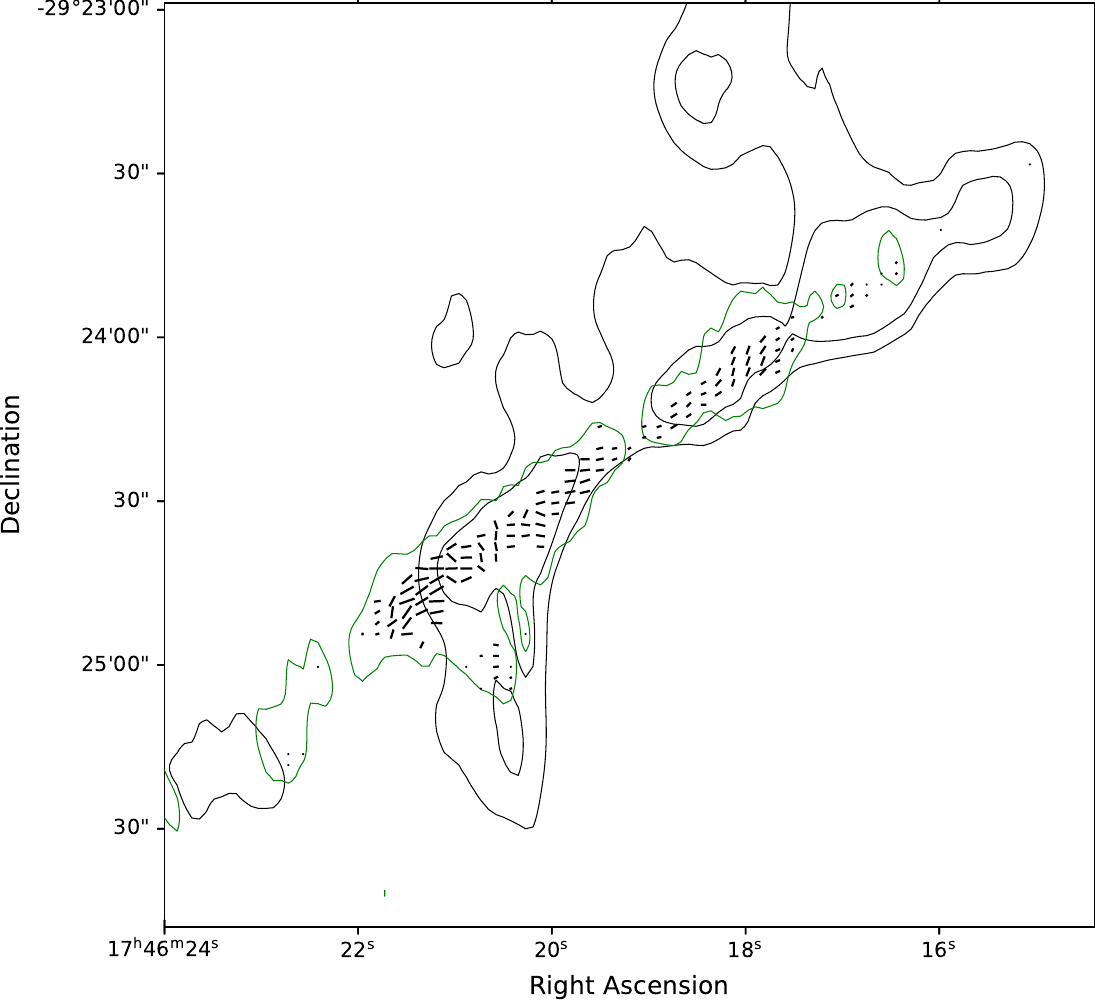}
    \caption{Intrinsic magnetic field orientations derived from the 10 GHz observations of SNTF3. Black contours indicate levels of constant total intensity at 0.25 and 0.35 mJy beam$^{-1}$, chosen to emphasize the total intensity structure discussed in the text. The green contour indicates a level of of constant polarized intensity at 0.1 mJy beam$^{-1}$ chosen to roughly indicate the significant polarized intensity of the filament.}
    \label{fig:SNTF3_X_B}
\end{figure*}

The intrinsic magnetic field orientations for SNTF3 from our 10 GHz observations are shown in Figure \ref{fig:SNTF3_X_B}. The magnetic field is almost uniformly parallel to the orientation of the brightest total intensity filament of SNTF3. This filament is the only one with significant debiased polarized intensity, and so we are unable to determine the magnetic field orientations in the fainter SNTF3 filaments.

In the Southern portion of SNTF3 at RA = 17:46:21, DEC = -29:24:48 the magnetic field becomes less uniform. Though the magnetic field is still generally parallel to the SNTF3 filament in this region, there are magnetic field orientations that are rotated by 60\degree\ relative to the filament orientation. This region coincides with the peak polarized intensity emission of the filament. It also coincides with a total intensity extension that diverges from the NTF filament length and extends to the South.

There is also a kink in the magnetic field orientation at RA = 17:46:17, DEC = -29:24:00. In this region the magnetic field changes from being parallel to the NTF filament to rotated by 30\degree\ to the NTF filament orientation. The location of this magnetic field kink coincides with a kink in the total intensity filament observed at 10 GHz \citep[Figure 10 of][]{Pare2022}. This kink in the magnetic field therefore means the magnetic field remains parallel to the local orientation of the SNTF1 filaments.

\section{DISCUSSION} \label{sec:disc}
\subsection{Differences in Polarized Intensity Features} \label{sec:extpol}
\subsubsection{Extended Polarized Intensity for SNTF1}
The majority of NTFs observed in the GC exhibit polarized intensity distributions that closely trace their total intensity emission \citep[e.g.,][]{YWP1997,Lang1999a,Lang1999b}. However, recent polarimetric observations of the prominent, multi-stranded Radio Arc NTFB revealed a polarized intensity distribution that extends into lower total intensity regimes \citep{Pare2019}. We see similar extensions for SNTF1 as shown in Figure \ref{fig:SNTF1P_2D}. Though the significant polarized intensity is generally confined within the SNTF1 structure as represented by the contour levels shown in Figure \ref{fig:SNTF1P_2D}, the polarized intensity extends between the individual filaments comprising the structure.

The extended polarized structures identified in \citet{Pare2019} were referred to as ``thorns,'' and could either be intrinsic to the Radio Arc or originate from some other polarized source along the line-of-sight. \citet{Pare2019} concluded that the thorns observed for the Radio Arc were associated with a shell of radio emission local to the Radio Arc filaments, but external to the NTFB. In the case of SNTF1 there is no apparent external structure that can be discerned from our VLA observations (like the Radio Shell local to the Radio Arc) that can readily explain the presence of the thorns in SNTF1.

However, we cannot rule out the possibility that there is an external structure larger than the largest angular size (LAS) to which our VLA observations are sensitive and therefore whose polarized intensity we are only partially resolving. The LAS of our VLA observations at 10 GHz is 240\arcsec\ ($\sim$9.5 pc for the GC). Our observations would therefore be insensitive to a structure that is contiguous on a scale larger than this angular size. However, the polarized intensity of a source is generally more discontinuous, as evidenced from previous polarimetric observations of GC structures at radio wavelengths \citep[e.g.,][]{Gray1995,Law2008}. We could therefore be recovering portions of the polarized distribution of this hypothetical external structure, even though we are not able to observe it in our total intensity distribution. This was the phenomenon that was occurring for the polarimetric observations of the Radio Arc that showed the thorn features of the shell, but could not resolve the Stokes $I$ distribution of the shell \citep{Pare2019}.

We attempt to identify a possible larger-scale, foreground structure that could be responsible for the thorns in SNTF1 by investigating single-dish observations. We do not observe an enhancement in the single-dish Nobeyama Radio Telescope (NRT) 10 GHz observations presented in \citet{Sofue1985}. However, more recent Green Bank Telescope (GBT) observations of Hn$\alpha$ RRL lines reveal a structure coincident with SNTF1 \citep{Anderson2024}. This structure also coincides with an \hii\ region labeled S20 in Figure 1 of \citet{Anderson2024} identified in the WISE Catalog of Galactic \hii\ regions \citep{Anderson2014}. This \hii\ region could be the foreground structure whose polarized intensity we are partially resolving. Follow-up, higher sensitivity observations are needed to verify whether this \hii\ region is the structure responsible for the polarized intensity thorns observed for SNTF1.

\subsubsection{Confined Polarized Intensity for SNTF3}
Conversely, the polarized intensity of SNTF3 (Figure \ref{fig:SNTF3_X_P_2D}) does not exhibit any of the polarized intensity extensions (or thorns) that are observed in SNTF1. This is a significantly different polarized intensity morphology than is seen for either SNTF1 or the Radio Arc. Furthermore, the polarized intensity distribution in SNTF3 parallels previous observations of single-stranded NTFs \citep[e.g.][]{Gray1995,Lang1999a,Lang1999b}.

There are multiple possibilities that could explain why the polarized intensity of SNTF3 is so distinct from that of SNTF1 and the Radio Arc, two of which are briefly discussed here. First, most of the SNTF3 filaments are quite faint compared to the noise level of the observations (SNR of $\sim$2). In fact, there is only one filament in SNTF3 that has a significant SNR relative to the noise level of the observations, and this is the filament that the polarized intensity traces as shown in Figure \ref{fig:SNTF3_X_P_2D}. It is therefore possible that the fainter filaments are polarized, but that we lack the sensitivity to resolve their polarized emission with the debiasing we employ. 

Alternatively, there is an apparent dependence on latitude observed in the polarized intensity distributions obtained from SNTF1 and SNTF3 presented here and the Radio Arc presented in \citet{Pare2019}. The Radio Arc is the closest to the Galactic plane (low latitudes) and exhibits the most prominent thorn features in its polarized intensity. SNTF3, which is the furthest in projection from the Galactic plane, exhibits no thorn features at all. SNTF1, which is at an intermediate latitude, has less pronounced thorns than the Radio Arc. Alternatively, the differences in thorn features might be simply a result of the presence or lack of intervening foreground media and disconnected from the difference in Galactic latitude of these NTFBs. 

\subsubsection{Differences in Fractional Polarization in the SNTFs}
The lower fractional polarizations in SNTF1 and SNTF3 could indicate a systematic difference between the NTFBs and the larger, single-stranded NTF population. However, the fractional polarizations for the Radio Arc ($\sim$0.4 \citep{Pare2019}) are in better agreement with the larger NTF population. The NTFBs therefore do not ubiquitously have lower fractional polarization than the larger, more isolated NTF population.

The "thorns" in both the Radio Arc and SNTF1 exhibit fractional polarizations $>$1.0, likely indicating that the total intensity associated with these extensions is connected to an external, magnetized structure with an angular size that is larger than the LAS of the VLA observations used to observe these NTFs \citep{Pare2019}, so the total flux values are undersampled in the thorn data, while the polarized flux values are less undersampled because they arise in generally smaller-scale features. This leads to the apparent fractional polarization appearing to be $>$1. Unlike with the Radio Arc, it is not as clear what the possible external candidate structure could be. It could, however, be a spatially coincident WISE \hii\ region \citep{Anderson2014}.

\subsection{The RM Distributions of the SNTFs}
\subsubsection{RMs of Different SNTFs}
The RM distribution for SNTF1 is generally in the range from 1000 to -1000 rad m$^{-2}$. There are regions within SNTF1 where the RM magnitude increases to $\rm|RM|>$1500 rad m$^{-2}$, but there is no systematic change in RM values along the length of the filaments. We also reiterate the point that the RM uncertainties in these spotty regions are large, meaning these RM magnitudes are largely consistent with $\rm|RM|<$1000 rad m$^{-2}$.

The relatively flat and uniform RM distribution of SNTF1 contrasts starkly with the RM distribution obtained for the Radio Arc \citep{Pare2019}. In the Radio Arc the RM was observed to systematically vary in magnitude throughout the length of the Radio Arc filaments. The RM variations were used in \citet{Pare2019} to infer the existence of a foreground magnetized medium present in some lines-of-sight towards the Radio Arc and not in others.

The comparatively flat and uniform nature of the SNTF1 RM distribution indicates that we are not encountering different numbers of foreground magnetized screens along the SNTF1 filament lengths. This lack of RM variation could indicate that there is no nearby GC-proximal medium impacting the RM distribution seen towards SNTF1. The lack of a nearby foreground medium could imply that the polarized emission obtained from SNTF1 shown in Figure \ref{fig:SNTF1P_2D} (i.e. the thorns) is predominantly associated with the SNTF1 filaments, rather than some other structure encountered along the line of sight. However, there could be a background structure located behind the SNTF1 filaments or an unresolved foreground structure that is contributing to the observed polarized intensity distribution. Furthermore, as discussed earlier, there is an \hii\ region identified in the GBT observations of \citet{Anderson2024} which could possibly be the foreground structure.

The SNTF3 RM distribution shown in Figure \ref{fig:SNTF3_X_RM} is also largely flat with magnitudes generally a few hundred rad m$^{-2}$. There is, however, a slight trend of increasing RM magnitudes in the southernmost declinations of SNTF3. In particular, the RM magnitudes observed at RA=17:45:22, DEC=-29.24.40 are generally $\rm|RM|>$600 rad m$^{-2}$. As mentioned previously, however, the RM values in this region should be viewed with caution since they have relatively larger uncertainties.

This region of elevated RM could indicate that these lines of sight are encountering a different number of magnetized media along the line of sight compared to the lines-of-sight throughout the length of the rest of the NTF. We do note that that this region of high RM coincides with an increase in polarized intensity, as can be seen in Figure \ref{fig:SNTF3_X_P_2D}. This region of elevated RM is also coincident with an extension in total intensity which can be seen in the contour levels of Figure \ref{fig:SNTF3_X_P_2D}. This extension is also observed in the 6 and 10 GHz observations of this NTFB presented in Figures 10 and 12 of \citet{Pare2022} where this is possibly a more diffuse, extended structure that crosses the SNTF3 filaments.

Through inspecting the NRT observations presented in \citet{Sofue1985} we identify an intensity enhancement that is spatially coincident with SNTF3. This enhancement is most visible in their Figure 3 at l, b = 359.7\degree, -0.3\degree. This detection does lend credence to the possibility that the total intensity structure observed in our 10 GHz VLA observations is a component of a larger structure. The more recent GBT Hn$\alpha$ RRL observations corroborate the existence of such a structure \citep{Anderson2024}. Furthermore, this structure coincides with an \hii\ region (feature S16 in their Figure 1) as identified in the WISE Catalog of Galactic \hii\ regions \citep{Anderson2014}.

\subsubsection{The GC-wide RM Distribution}
\begin{figure*}
    \centering
    \includegraphics[width=1.0\textwidth]{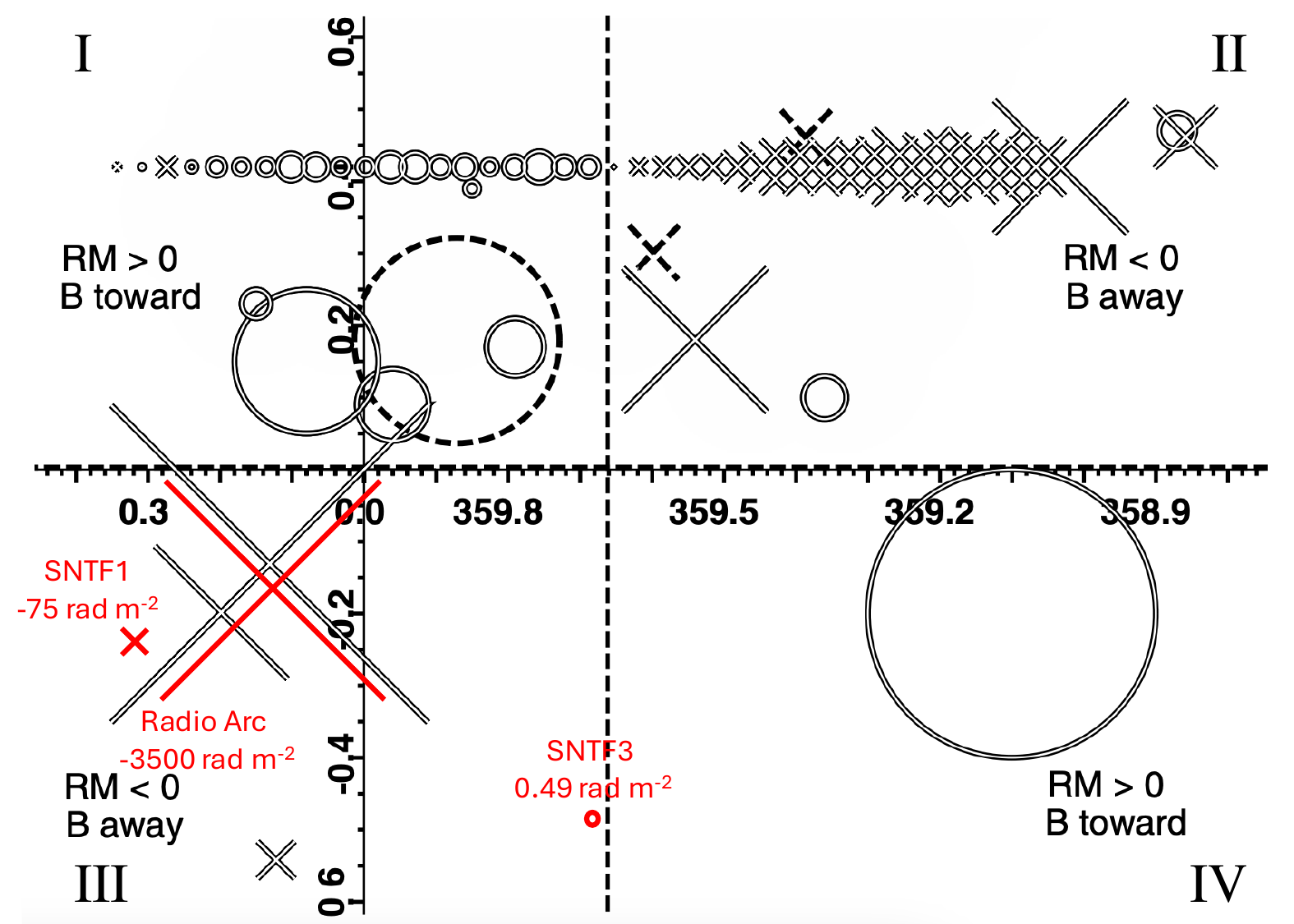}
    \caption{The large-scale RM distribution observed towards the GC. In this figure the extent of the GC in the plane of the sky has been split into four quadrants (labeled I -- IV in the figure). Circles represent positive RM values and crosses represent negative RM values, with the size of each symbol scaled by RM magnitude. Black circles and crosses represent the RM values collated previously by \citet{Law2011}. Red circles and crosses represent new RM values studied in this and other recent radio polarimetric studies of the GC \citep{Pare2019,Pare2021}.}
    \label{fig:GC_RM}
\end{figure*}
The new RM distributions obtained from SNTF1 and SNTF3 are also useful for characterizing large-scale magnetic field orientations in the GC. Previous efforts have been made to characterize the large-scale magnetic field observed throughout the GC. \citet{Law2011}, for example, collated existing radio polarimetric results obtained for the GC and found a checkerboard pattern in the large-scale RM distribution towards the GC. Though comprehensive, this effort was limited by the paucity of polarimetric observations towards the GC, particularly at Southern Galactic latitudes below the Galactic plane where not many polarimetric observations had been reported prior to that time.

We build on the work presented in \citet{Law2011} to characterize the large-scale RM distribution obtained toward the GC by incorporating the polarimetric results presented in this work and other recent radio polarimetric studies of the GC \citep[e.g.,][]{Pare2019,Pare2021}. The result of this updated GC-wide analysis is shown in Figure \ref{fig:GC_RM}. In this figure, the previous \citet{Law2011} results are shown as black circles and crosses and the more recent observations are shown as red circles and crosses.

The new RM measurements that have been observed towards the GC are consistent with the checkerboard pattern suggested by \citet{Law2011}. This checkerboard pattern is shifted to the West of the Galactic Center based on the conclusion of \citet{Law2011} that the pattern more robustly fits the RM measurements with this shift. Quadrants I and IV have largely positive RM values, with quadrants II and III having largely negative RM values. Negative RM values indicate a magnetic field that is oriented away from the observer, whereas positive RM values indicate a magnetic field that is oriented towards the observer, along the line of sight.

Even prior to the work of \citet{Law2011} this checkerboard RM pattern had been suggested \citep{Uchida1985,Novak2003a}. This checkerboard pattern motivated the theory proposed by \citet{Uchida1985} and \citet{Novak2003a} of magnetic flux dragging caused by the orbital rotation in the GC. This theory posits that the GC  originally experienced a pervasive poloidal (or vertical) field which was then deformed by flux-freezing of the magnetic field into the differentially rotating gas layer of the CMZ. The first perturbation on this distribution would result in a checkerboard pattern in the line-of-sight magnetic field.

Observations in Southern latitudes are still sparse compared to the Northern quadrants. However, the fact that these more recent RM observations corroborate the checkerboard pattern of \citet{Law2011} strengthens the possibility that the magnetic flux is being dragged by the orbital motion of the GC. The RM point associated with SNTF3 has a positive RM value and is located within the third quadrant, which otherwise exhibits negative RM values. However, we note that the average RM value of this NTFB is very close to 0, and so this point could mark the transition region between quadrant III and IV. This low RM value therefore also corroborates the checkerbord RM pattern. 

The possibility that the magnetic field is drgged by the orbital motion of the GC is also corroborated by recent far-infrared dust polarization observations that show evidence that a poloidal (or vertical) magnetic field evidenced by NTFs throughout the GC has been sheared to trace the orientation of the molecular clouds in the GC \citep{Pare2024}. Further observations of Southern-latitude polarized radio sources are needed to assess the checkerboard pattern more robustly. We note that the RM values presented in Figure \ref{fig:GC_RM} do not include the RM values found for magnetars, pulsars, and Sgr A$^*$ \citep[e.g.,][]{Desvignes2018,Abbate2023}, since these sources typically have strong RM values ($\geq$10,000 rad m$^{-2}$) not representative of the values seen in the larger GC region or in the NTF population (with values of $\sim$1000 rad m$^{-2}$).

\subsection{The Intrinsic Magnetic Fields of the SNTFs}
Previous polarimetric observations of the Radio Arc reveal an alternating magnetic field pattern where the field regularly switches from being parallel to rotated with respect to the Radio Arc filament orientation \citep{Pare2019}. The larger NTF population, conversely, exhibits generally uniform magnetic fields oriented parallel to the NTF filament orientations \citep[e.g.,][]{Gray1995,Lang1999a,Lang1999b}. 

The intrinsic magnetic field derived for SNTF1 (Figure \ref{fig:SNTF1_X_B}) is spatially variant. We note an alternating magnetic field pattern along the SNTF1 filament that is similar to what was previously observed toward the Radio Arc \citep{Pare2019}, although this pattern is less well ordered than that observed for the Radio Arc. The increased spatial variance of the SNTF1 intrinsic magnetic field could be partially attributed to the fainter emission of SNTF1 relative to the Radio Arc. In addition to this pattern, there is also a slight trend of more parallel magnetic field in the Northern SNTF1 filaments than in the Southern SNTF1 filaments.

The trend from North to South in the magnetic field of SNTF1 could provide important insight into the geometry of the magnetized screens encountered along the line-of-sight. The perpendicular magnetic field could be originating from a foreground magnetized screen, since our linear model for Faraday correction may not fully correct for foreground Faraday effects if there are changing numbers of intervening magnetized media. The parallel magnetic field regions could be windows through this screen that reveal the intrinsic magnetic field of the SNTF1 filaments.

The generally parallel magnetic field of SNTF3 is in good agreement with the majority of magnetic field observations obtained from previous GC NTF observations. There is no significant evidence of an alternating magnetic field pattern like that observed in SNTF1 and the Radio Arc \citep{Pare2019}. In fact, there are no significant magnetic field orientation changes throughout the SNTF3 filament length.

Furthermore, the fact that the kink in the magnetic field coincides with a kink observed in the 10 GHz total intensity distribution for SNTF3 \citep{Pare2022} demonstrates that the polarization and magnetic field derived from these observations originates from the SNTF3 structure rather than from some other magnetized medium along the line-of-sight.

\subsection{Evaluating Possible Relativistic Electron Acceleration Mechanisms} \label{sec:theory}
\citet{Pare2022} leveraged the total intensity and spectral index properties of the NTFBs to propose possibilities for how the relativistic electrons illuminating these structures are generated. The conclusion from \citet{Pare2022} was that SNTF1 was likely illuminated by CRs generated by a nearby pulsar or some other compact source, and those electrons then diffuse and propagate along the NTF filaments. Conversely, SNTF3 was suggested to likely be illuminated by either multiple compact sources of CRs or by an extended source of CRs.

In this section we build on these initial conclusions of \citet{Pare2022} by incorporating the polarimetric, RM, and intrinsic magnetic field distributions presented in this work to expand on the possibilities for how the relativistic electrons illuminating these NTFs could have been generated.

\subsubsection{SNTF1 Illuminated by CRs From a Compact Source}
The total intensity morphology of SNTF1 implies that the Northern filaments are closer to the compact source of CRs (since they are shorter and brighter) than the Southern filaments. \citet{Pare2022} argued that the total intensity morphology of SNTF1 paralleled that of the harps studied by \citet{Thomas2020}, where the possible compact source of CRs could be the point source with a jet or tail of emission identified by \citet{Pare2022}. 

The polarized intensity distribution derived for SNTF1 in this work (shown in Figure \ref{fig:SNTF1P_2D}) corroborates the theory that the filaments in SNTF1 are illuminated by CRs originating from a compact source located to the North of SNTF1. The polarimetric distribution for SNTF1 exhibits more significant polarization associated with the Northern regions of the SNTF1 filaments. In more Southern regions in SNTF1 the significant polarized intensity becomes more sparse and discontinuous. Furthermore, the polarized intensity magnitudes are generally lower in the Southern portion of SNTF1.

The more significant polarized intensity in the Northern portion of SNTF1 indicates the presence of a larger number of relativistic electrons propagating along these filaments relative to the filaments in the Southern portion of SNTF1. The Northern filaments are also brighter in total intensity. This morphology corroborates the idea that the Northern filaments are closer to a compact source of CRs and therefore the electrons illuminating these filaments have had less time to diffuse and propagate along the flux tubes traced by these filaments. This assumes that the compact source is moving laterally to the field direction and injects CRs into the magnetic flux tubes traced by the NTFs as the source crosses the filaments.

An alternative explanation for the more sparse polarized intensity in the Southern filaments could simply be a result of these filaments being fainter and having a lower SNR relative to the noise level of our VLA observations. We cannot rule out this alternative explanation, and follow-up polarimetric observations with a greater sensitivity are needed to verify how the polarization and total intensity distributions vary between the Northern and Southern NTF filaments. We note that in the VLA observations presented here we observe no significant fractional polarization change.

We see no significant trends in the SNTF1 RM distribution in Figure \ref{fig:SNTF1_RM} when comparing the RM magnitudes observed in the Northern and Southern SNTF1 filaments. However, since the RM distribution is predominantly sensitive to Faraday effects in foreground magnetized screens, we would not necessarily expect there to be any such systematic differences.

The intrinsic magnetic field of SNTF1 in Figure \ref{fig:SNTF1_X_B} shows more parallel magnetic field orientations associated with the Northern SNTF1 filaments compared to the Southern filaments. The change in apparent magnetic field orientation could indicate a more complex foreground medium encountered by the lines-of-sight in the Southern portion of SNTF1 relative to the North. \citet{Pare2019} inferred the existence of different numbers of magnetized media from the alternating magnetic field pattern they observed towards the Radio Arc. The similar alternating pattern of SNTF1 could also indicate the presence of changing intervening magnetized media along the different lines of sight towards SNTF1. A more detailed modeling of the intervening Faraday effects encountered towards SNTF1, coupled with higher resolution observations possible with next-generation instruments like the ngVLA, is needed to evaluate this possibility.

\subsubsection{Implications from SNTF3 Polarimetric Distributions}
We only see significant polarized intensity associated with the brightest filament in SNTF3. The magnitude of the polarized intensity steadily decreases from Southern declinations to Northern declinations, as seen in Figure \ref{fig:SNTF3_X_P_2D}. \citet{Pare2022} concluded that SNTF3 was likely illuminated by relativistic electrons emanating from either an extended source or multiple compact sources. This conclusion was based on the fact that the filaments in SNTF3 were of similar length and there is no obvious total intensity brightness gradient across the SNTF3 structure. The more significant polarized intensity in Southern declinations could indicate that this portion of the NTF filaments is where the relativistic electrons illuminating these structures are injected into the SNTF3 structure.

This conclusion is corroborated by the spectral index distribution derived for SNTF3 by  \citet{Pare2022}. They found a steeper spectral index at lower Galactic latitudes that progressed to more shallow spectral indices at higher Galactic latitudes. We note, however, that the uncertainties on the spectral index measurements for SNTF3 shown in Figure 13 of \citet{Pare2022} reduce the certainty of that conclusion. 

The region of peak polarized intensity from the polarimetric results presented in this paper is the Southernmost polarized portion of SNTF3. This is the portion of the polarized distribution that is closest in angular separation to the steeper spectral index regions of SNTF3 observed in \citet{Pare2022}. The combination of polarized intensity and spectral index results therefore possibly indicates that the Southern extent of the SNTF3 structure is closer to the source of relativistic electrons illuminating the structure. 

Incorporating the detection of a large-scale structure in the NRT and GBT observations presented in \citet{Sofue1985} and \citet{Anderson2024}, we prefer the possibility of a single, extended source of CRs being responsible for generating the relativistic electrons illuminating this structure. Additional work is needed to determine the nature of this extended source.

\section{CONCLUSIONS} \label{sec:conc}
In this paper we study the polarimetric distributions associated with a set of GC NTFBs. This is a follow-up study of the same NTF structures observed and discussed in \citet{Pare2022}. We obtain significant polarized intensity for SNTF1 and SNTF3 at 10 GHZ, and for the Wishbone at 6 GHz. Using these polarized intensities we derived the RM and intrinsic magnetic field distributions for these targets. We then discuss the possible electron generation mechanisms that could explain how the target NTFBs are illuminated, building on the discussion provided by \citet{Pare2022}.

We summarize the key takeaways of this paper here:
\begin{itemize}
    \item The polarized intensity distribution of SNTF1 exhibits extensions that are not associated with significant total intensity emission. These extensions are similar to what was previously observed for the Radio Arc  \citep{Pare2019}. Conversely, the polarized intensity distributions of SNTF3 and the Wishbone closely trace the total intensity distributions of these structures.
    \item The magnetic field is spatially variant in SNTF1: regions where the magnetic field is oriented parallel to the SNTF1 filaments are separated by a region where the magnetic field is rotated or perpendicular to the filaments. SNTF3, conversely, exhibits a generally parallel intrinsic magnetic field. The alternating magnetic field pattern of SNTF1 could indicate the presence of a magnetized foreground structure that is leaving an imprint on the polarized intensity. This structure could possibly be an \hii\ region, as identified in the GBT observations presented in \citet{Anderson2024}.
    \item We expand on previous efforts to model the RM and magnetic field distribution on GC-wide scales by collating previous RM observations towards the GC. Our results with these new observations corroborate the checkerboard RM pattern observed previously in \citet{Law2011}. This pattern is indicative of a magnetic field which has become perturbed by magnetic flux dragging caused by orbital motion in the region. This GC-wide field supports previous theoretical work and recent observational results that indicate an initially vertical field became sheared in the higher density GC molecular clouds due to the orbital motion of these clouds \citep{Uchida1985,Chuss2003a,Pare2024}.
    \item Finally, we have used our polarimetric results to expand on the work presented in \citet{Pare2022} to determine how the relativistic electrons illuminating the NTFs are generated. We find that our polarimetric distributions corroborate the conclusions of \citet{Pare2022} that SNTF1 is likely illuminated by a compact source of CRs. We determine that SNTF3 is likely illuminated by an extended source of CRs rather than multiple compact sources, with the extended source possibly being an \hii\ region.
\end{itemize}

\acknowledgements We would like to thank the anonymous referee for their helpful comments and suggestions on the paper. The National Radio Astronomy Observatory is a facility of the National Science Foundation operated under cooperative agreement by Associated Universities, Inc. This work was supported in part by National Science Foundation grants AST-1614782 to UCLA and AST-1615375 to the University of Iowa.

\software{
    MIRIAD, \citep{Sault1995},
    Astropy \citep{Greenfield2014},
    CASA \citep{McMullin2007},
    Matplotlib \citep{Hunter2007}
    }

\bibliographystyle{aasjournal}
\bibliography{astronomy}

\end{document}